\documentclass[twocolumn,preprintnumbers,amsmath,amssymb]{revtex4}
\newcommand{\eps}{\varepsilon}

\newcommand{\vct}[1]{\mbox{\boldmath #1}}
\catcode`\@=11
\def\eqlt{\mathrel{\mathpalette\@vereq<}}  
\def\eqgt{\mathrel{\mathpalette\@vereq>}}  
\def\@vereq#1#2{\lower2.5pt\vbox{\baselineskip0pt \lineskip-.5pt
 \ialign{$\m@th#1\hfil##\hfil$\crcr#2\crcr{=}\crcr}}}

\newcommand{\simle}{\ \raise.3ex\hbox{$<$}\kern-0.8em\lower.7ex\hbox{$\sim$}\ }
\newcommand{\simge}{\ \raise.3ex\hbox{$>$}\kern-0.8em\lower.7ex\hbox{$\sim$}\ }

\begin{document}
\title {Orbital-Spin Structure and Coupling to Lattice in $R$TiO$_3$ with $R=$La,Pr,Nd and Sm}  
\author {  Masahito Mochizuki and Masatoshi Imada }  
\address {Institute for Solid State Physics, University of Tokyo, Kashiwanoha,
Kashiwa, Chiba, 277-8581, Japan}  
\date { December 6, 2002 }  

\begin{abstract} 
The origin of the G-type antiferromagnetism (AFM(G)) and 
puzzling properties of RTiO$_3$ with $R$=La are studied.
We clarify that the crystal field from La  
caused by the GdFeO$_3$-type distortion lifts the $t_{2g}$ degeneracy at Ti $3d$ orbitals.
The lift stabilizes the AFM(G) with 
spin-exchange constant in agreement with neutron scattering results. 
The orbital-spin structures for
$R$=Pr, Nd and Sm are also consistent with experiments.
We propose that the GdFeO$_3$-type distortion has a universal mechanism of controlling 
orbital-spin structure competing with the Jahn-Teller (JT) mechanism.\end{abstract}

\maketitle
Rich magnetic and orbital phases, their phase transitions
and interplays have attracted much interest in strongly 
correlated electron systems, particularly in transition metal (TM) oxides. 
Coupling to lattice further enriches the interplay through lattice
distortions, dynamical phonons and cooperative effects such as JT 
distortions~\cite{Imada98}. Cuprate superconductors and manganese 
perovskite compounds with colossal magnetoresistance belong to 
the TM oxides with $3d$ $e_g$ bands at the Fermi level.
Perovskite-type Ti oxides $R$TiO$_3$ with 
$R$ being a rare-earth ion belong to the same type but with 
$3d$ $t_{2g}$ bands at the Fermi level and show a very different behavior.
Particularly, while the above mother materials 
are commonly typical Mott insulators, $R$TiO$_3$ shows markedly 
rich and complicated behaviors possibly due to the above interplay.
The crystal structure of $R$TiO$_3$ is a pseudocubic perovskite with an orthorhombic 
distortion (GdFeO$_3$-type distortion) and exhibits a magnetic 
phase transition as a function of the magnitude of this 
distortion~\cite{Okimoto95,Katsufuji97,Goral82,Greedan85}. 
The end compounds $R=$Gd and Y with large GdFeO$_3$-type distortions
showing the ferromagnetic (FM) ground state
are relatively well understood~\cite{Mochizuki00,Mochizuki01a}.
In contrast, the magnetic structure of LaTiO$_3$ with relatively small
GdFeO$_3$-type distortion has AFM(G) structure with N$\rm {\acute{e}}$el 
temperature ($T_N$) of $\sim130$ K~\cite{Goral82}. The origin of this G-type 
antiferromagnetism has long been puzzling and controversial.
The purpose of the present theoretical study is to solve this puzzle
and offer an overall understanding of the interplay in these compounds.

In perovskite TM compounds, the
JT distortions often play crucial roles in determining low-energy
electronic states. 
However, with early TM ions, the 
JT coupling is considerably weaker than that in the late-$3d$ compounds.
Actually, the distortion of the TiO$_6$ octahedra in LaTiO$_3$ has not 
been detected in contrast to the manganites and cuprates~\cite{MacLean79,Eitel86}.
Therefore, in LaTiO$_3$, the crystal field of 
O ions surrounding the ${\rm Ti}^{3+}$ ion has a cubic symmetry so that the 
degeneracy of the $t_{2g}$ level survives. 
Under this circumstance, the AFM(G) ground state is 
surprising and controversial since in the orbitally degenerate system,
we expect that a FM state with antiferro-orbital ordering is stabilized both 
by transfers and the Coulomb exchange interaction~\cite{Imada98}. 
Indeed, a recent weak coupling study shows that a FM state, out of which two states 
$(yz+izx)/\sqrt{2}\uparrow$ and $xy\uparrow$
are alternating is favored both by the relativistic spin-orbit (LS) interaction and by the 
spin-orbital superexchange interaction~\cite{Mochizuki02}.
 
Moreover, a recent neutron-scattering 
study shows the spin-wave spectrum well described by an isotropic 
spin-1/2 Heisenberg model with a nearest-neighbor superexchange 
constant $J\sim15.5$ meV~\cite{Keimer00}. 
If the orbital moment is nonzero, the anisotropy appears through the LS interaction.
Therefore, the isotropy indicates that the orbital moments are quenched.

Recently, a possible orbital liquid state was proposed by assuming the G-type
antiferromagnetism on the 
basis of resultant small exchange interaction in the orbital sector~\cite{Khaliullin00}.
However, the origin of the AFM(G) state in LaTiO$_3$ is still puzzling when the 
orbital is disordered.
On the other hand, the possibility of the $D_{3d}$ distortion of the TiO$_6$ 
octahedra was also proposed for the origin of 
the AFM(G)~\cite{Mochizuki01b}.
In this distortion, the TiO$_6$ octahedron is contracted along the 
threefold direction, and the threefold-degenerate $t_{2g}$-levels split 
into a nondegenerate lower $a_{1g}$-level and twofold-degenerate higher 
$e_g$-levels. The lift of the $t_{2g}$ levels with the occupation of the 
lowest $a_{1g}$-orbital well explains the emergence of the 
AFM(G) and the isotropic spin-wave spectrum.
However, this $D_{3d}$ distortion has so far not been 
detected.

In addition to the JT distortion, the GdFeO$_3$-type lattice 
distortion is another generic phenomenon in perovskites. 
It has been assumed mainly to control the bandwidth 
through the $M$-O-$M$ angle with $M$ being TM ions, while its direct 
effects on interplay of spins and orbitals have not been considered 
seriously. 

In this letter, by taking a particular example of LaTiO$_3$,
we clarify that the generic GdFeO$_3$-type distortion actually generates a 
new mechanism for control of orbital-spin low-energy structure through
lift of the orbital degeneracy by crystal fields of $R$ ions.
This mechanism competes with that of the JT distortions and LS interactions.
By utilizing the experimentally obtained 
coordination parameters, we construct the Hamiltonian for the crystal 
field of La cations in LaTiO$_3$. 
The analyses of the obtained Hamiltonian show that the shifts of La cations 
due to the GdFeO$_3$-type distortion turn out to generate the crystal field 
which is similar to the $D_{3d}$-crystal-field.
As a result, the threefold degeneracy of the cubic-$t_{2g}$-levels splits 
into three nondegenerate levels. The calculations of the energies and 
the spin-exchange constant well explain the emergence and the properties of the 
AFM(G) state in LaTiO$_3$. 
In addition, we also study the effects of the crystal field of rare-earth 
ions in $R$TiO$_3$ with $R$ being Pr, Nd and Sm.
Again experimentally observed orbital structure as well as the reduction of spin exchange are reproduced.
We point out the importance of GdFeO$_3$-type 
distortion as a generic control mechanism of orbital-spin structure in perovskite
compounds.

%
%
%
%
%
\catcode`\@=11
%
%
\def\psfortextures{
\def\PSspeci@l##1##2{%
\special{illustration ##1\space scaled ##2}%
}}
\def\psfordvitops{
\def\PSspeci@l##1##2{%
\special{dvitops: import ##1\space \the\drawingwd \the\drawinght}%
}}
\def\psfordvips{
\def\PSspeci@l##1##2{%
\d@my=0.1bp \d@mx=\drawingwd \divide\d@mx by\d@my%
\includegraphics{##1\space}%
}}
\def\psforoztex{
\def\PSspeci@l##1##2{%
\special{##1 \space
      ##2 1000 div dup scale
      \putsp@ce{\number-\psllx} \putsp@ce{\number-\pslly} translate
}%
}}
\def\putsp@ce#1{#1 }
\def\psfordvitps{
\def\psdimt@n@sp##1{\d@mx=##1\relax\edef\psn@sp{\number\d@mx}}
\def\PSspeci@l##1##2{%
\special{dvitps: Include0 "psfig.psr"}
\psdimt@n@sp{\drawingwd}
\special{dvitps: Literal "\psn@sp\space"}
\psdimt@n@sp{\drawinght}
\special{dvitps: Literal "\psn@sp\space"}
\psdimt@n@sp{\psllx bp}
\special{dvitps: Literal "\psn@sp\space"}
\psdimt@n@sp{\pslly bp}
\special{dvitps: Literal "\psn@sp\space"}
\psdimt@n@sp{\psurx bp}
\special{dvitps: Literal "\psn@sp\space"}
\psdimt@n@sp{\psury bp}
\special{dvitps: Literal "\psn@sp\space startTexFig\space"}
\special{dvitps: Include1 "##1"}
\special{dvitps: Literal "endTexFig\space"}
}}
\def\psonlyboxes{
\def\PSspeci@l##1##2{%
\at(0cm;0cm){\boxit{\vbox to\drawinght
  {\vss
  \hbox to\drawingwd{\at(0cm;0cm){\hbox{(##1)}}\hss}
  }}}
}%
}
\def\psloc@lerr#1{%
\let\savedPSspeci@l=\PSspeci@l%
\def\PSspeci@l##1##2{%
\at(0cm;0cm){\boxit{\vbox to\drawinght
  {\vss
  \hbox to\drawingwd{\at(0cm;0cm){\hbox{(##1) #1}}\hss}
  }}}
\let\PSspeci@l=\savedPSspeci@l
}%
}
%
%
\newread\psiz@
\newdimen\drawinght\newdimen\drawingwd
\newdimen\psxoffset\newdimen\psyoffset
\newbox\drawingBox
\newif\ifNotB@undingBox
\newhelp\PShelp{Proceed: you'll have a 5cm square blank box instead of
your graphics (Jean Orloff).}
\def\@mpty{}
\def\s@tsize#1 #2 #3 #4\@ndsize{
  \def\psllx{#1}\def\pslly{#2}%
  \def\psurx{#3}\def\psury{#4}
  \ifx\psurx\@mpty\NotB@undingBoxtrue
  \else
    \drawinght=#4bp\advance\drawinght by-#2bp
    \drawingwd=#3bp\advance\drawingwd by-#1bp
  \fi
  }
\def\sc@nline#1:#2\@ndline{\edef\p@rameter{#1}\edef\v@lue{#2}}
\def\g@bblefirstblank#1#2:{\ifx#1 \else#1\fi#2}
\def\psm@keother#1{\catcode`#112\relax}
\def\execute#1{#1}
{\catcode`\%=12
\xdef\B@undingBox{
}  		
\def\ReadPSize#1{
 \edef\PSfilename{#1}
 \openin\psiz@=#1\relax
 \ifeof\psiz@ \errhelp=\PShelp
   \errmessage{I haven't found your postscript file (\PSfilename)}
   \psloc@lerr{was not found}
   \s@tsize 0 0 142 142\@ndsize
   \closein\psiz@
 \else
   \loop
     \execute{\begingroup
       \let\do\psm@keother
       \dospecials
       \catcode`\ =10
       \catcode`\^^M=9
       \global\read\psiz@ to\n@xtline
       \endgroup}
     \ifeof\psiz@
       \errhelp=\PShelp
       \errmessage{(\PSfilename) is not an Encapsulated PostScript File:
           I could not find any \B@undingBox: line.}
       \edef\v@lue{0 0 142 142:}
       \psloc@lerr{is not an EPSFile}
       \NotB@undingBoxfalse
     \else
       \expandafter\sc@nline\n@xtline:\@ndline
       \ifx\p@rameter\B@undingBox\NotB@undingBoxfalse
         \edef\int@rmediateresult{%
           \expandafter\g@bblefirstblank\v@lue\space\space\space}
         \expandafter\s@tsize\int@rmediateresult\@ndsize
       \else\NotB@undingBoxtrue
       \fi
     \fi
   \ifNotB@undingBox\repeat
   \closein\psiz@
 \fi
\message{#1}
}
%
%
\newcount\xscale \newcount\yscale \newdimen\pscm\pscm=1cm
\newdimen\d@mx \newdimen\d@my
\let\ps@nnotation=\relax
\def\psboxto(#1;#2)#3{\vbox{
   \ReadPSize{#3}
   \divide\drawingwd by 1000
   \divide\drawinght by 1000
   \d@mx=#1
   \ifdim\d@mx=0pt\xscale=1000
         \else \xscale=\d@mx \divide \xscale by \drawingwd\fi
   \d@my=#2
   \ifdim\d@my=0pt\yscale=1000
         \else \yscale=\d@my \divide \yscale by \drawinght\fi
   \ifnum\yscale=1000
         \else\ifnum\xscale=1000\xscale=\yscale
                    \else\ifnum\yscale<\xscale\xscale=\yscale\fi
              \fi
   \fi
   \divide \psxoffset by 1000\multiply\psxoffset by \xscale
   \divide \psyoffset by 1000\multiply\psyoffset by \xscale
   \global\divide\pscm by 1000
   \global\multiply\pscm by\xscale
   \multiply\drawingwd by\xscale \multiply\drawinght by\xscale
   \ifdim\d@mx=0pt\d@mx=\drawingwd\fi
   \ifdim\d@my=0pt\d@my=\drawinght\fi
   \message{scaled \the\xscale}
 \hbox to\d@mx{\hss\vbox to\d@my{\vss
   \global\setbox\drawingBox=\hbox to 0pt{\kern\psxoffset\vbox to 0pt{
      \kern-\psyoffset
      \PSspeci@l{\PSfilename}{\the\xscale}
      \vss}\hss\ps@nnotation}
   \global\ht\drawingBox=\the\drawinght
   \global\wd\drawingBox=\the\drawingwd
   \baselineskip=0pt
   \copy\drawingBox
 \vss}\hss}
  \global\psxoffset=0pt
  \global\psyoffset=0pt
  \global\pscm=1cm
  \global\drawingwd=\drawingwd
  \global\drawinght=\drawinght
}}
%
%
\def\psboxscaled#1#2{\vbox{
  \ReadPSize{#2}
  \xscale=#1
  \message{scaled \the\xscale}
  \divide\drawingwd by 1000\multiply\drawingwd by\xscale
  \divide\drawinght by 1000\multiply\drawinght by\xscale
  \divide \psxoffset by 1000\multiply\psxoffset by \xscale
  \divide \psyoffset by 1000\multiply\psyoffset by \xscale
  \global\divide\pscm by 1000
  \global\multiply\pscm by\xscale
  \global\setbox\drawingBox=\hbox to 0pt{\kern\psxoffset\vbox to 0pt{
     \kern-\psyoffset
     \PSspeci@l{\PSfilename}{\the\xscale}
     \vss}\hss\ps@nnotation}
  \global\ht\drawingBox=\the\drawinght
  \global\wd\drawingBox=\the\drawingwd
  \baselineskip=0pt
  \copy\drawingBox
  \global\psxoffset=0pt
  \global\psyoffset=0pt
  \global\pscm=1cm
  \global\drawingwd=\drawingwd
  \global\drawinght=\drawinght
}}
%
\def\psbox#1{\psboxscaled{1000}{#1}}
%
%
\def\centinsert#1{\midinsert\line{\hss#1\hss}\endinsert}
\def\psannotate#1#2{\def\ps@nnotation{#2\global\let\ps@nnotation=\relax}#1}
\def\pscaption#1#2{\vbox{
   \setbox\drawingBox=#1
   \copy\drawingBox
   \vskip\baselineskip
   \vbox{\hsize=\wd\drawingBox\setbox0=\hbox{#2}
     \ifdim\wd0>\hsize
       \noindent\unhbox0\tolerance=5000
    \else\centerline{\box0}
    \fi
}}}
\def\psfig#1#2#3{\pscaption{\psannotate{#1}{#2}}{#3}}
\def\psfigurebox#1#2#3{\pscaption{\psannotate{\psbox{#1}}{#2}}{#3}}
%
\def\at(#1;#2)#3{\setbox0=\hbox{#3}\ht0=0pt\dp0=0pt
  \rlap{\kern#1\vbox to0pt{\kern-#2\box0\vss}}}
%
\newdimen\gridht \newdimen\gridwd
\def\gridfill(#1;#2){
  \setbox0=\hbox to 1\pscm
  {\vrule height1\pscm width.4pt\leaders\hrule\hfill}
  \gridht=#1
  \divide\gridht by \ht0
  \multiply\gridht by \ht0
  \gridwd=#2
  \divide\gridwd by \wd0
  \multiply\gridwd by \wd0
  \advance \gridwd by \wd0
  \vbox to \gridht{\leaders\hbox to\gridwd{\leaders\box0\hfill}\vfill}}
%
\def\fillinggrid{\at(0cm;0cm){\vbox{
  \gridfill(\ht\drawingBox;\wd\drawingBox)}}}
%
%
\def\textleftof#1:{
  \setbox1=#1
  \setbox0=\vbox\bgroup
    \advance\hsize by -\wd1 \advance\hsize by -2em}
\def\textrightof#1:{
  \setbox0=#1
  \setbox1=\vbox\bgroup
    \advance\hsize by -\wd0 \advance\hsize by -2em}
\def\endtext{
  \egroup
  \hbox to \hsize{\valign{\vfil##\vfil\cr%
\box0\cr%
\noalign{\hss}\box1\cr}}}
%
\def\frameit#1#2#3{\hbox{\vrule width#1\vbox{
  \hrule height#1\vskip#2\hbox{\hskip#2\vbox{#3}\hskip#2}%
        \vskip#2\hrule height#1}\vrule width#1}}
\def\boxit#1{\frameit{0.4pt}{0pt}{#1}}
\catcode`\@=12 
%
 \psfordvips   

\begin{figure}
$$ \psboxscaled{400}{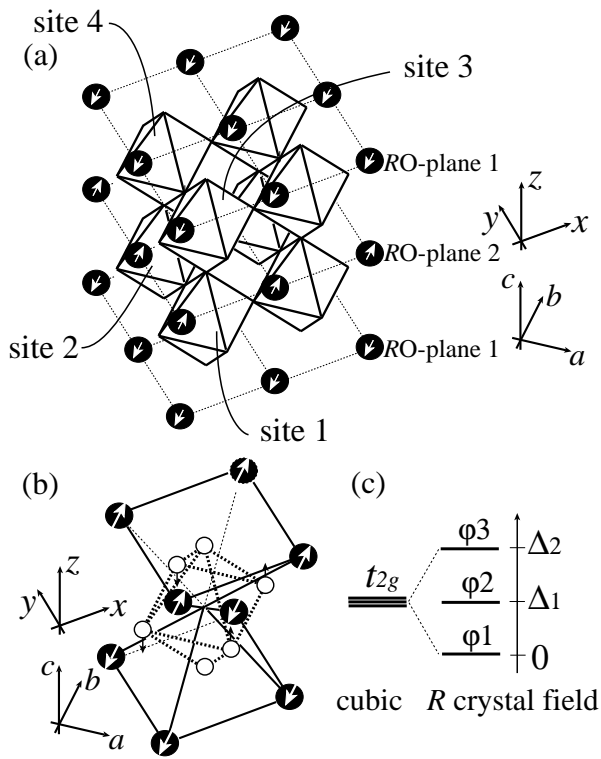} $$
\caption{(a) The distortions of the $R$ cations along the $b$-axis due
to the GdFeO$_3$-type distortion are presented. In $R$O-plane 1, they 
distort in the positive direction while in $R$O-plane 2, they distort in 
the negative direction. (b) In the case of site 1, the distances 
between the Ti ion and $R$ ions located in the $\pm(1,1,-1)$-directions 
are decreased. As a result, the La ions generate the crystal field 
which is similar to the $D_{3d}$-crystal field with 
$[1,1,-1]$-trigonal axis (see text). 
The distortions of the O ions along the $c$-axis which are induced 
by the tilting of the octahedron are also presented. (c) The 
energy-level structure in the crystal field due to the $R$
cations. The threefold degeneracy of the cubic-$t_{2g}$ levels 
split into three nondegenerate levels.}
\label{reshift}
\end{figure}
In the GdFeO$_3$-type distortion, the $R$ ions distort mainly along 
the $(1,1,0)$-axis or the $b$-axis, and slightly along the $(1,-1,0)$-axis 
or the $a$-axis. In Fig.~\ref{reshift} (a), we show the distortions along 
the $b$-axis. There are two kinds of $R$O-planes (plane 1 and plane 2) 
stacking alternatingly along the $c$-axis. In plane 1, the $R$ ions 
shift in the positive direction while they shift in the negative 
direction in plane 2. Consequently, the crystal field from the 
$R$ cations are distorted from a cubic symmetry.
For example, in the case of sites 1 and 2, the distances between the Ti 
ion and the $R$ ions located in the $\pm(1,1,-1)$-directions decrease 
while those along the $\pm(1,1,1)$-directions increase 
(see also Fig.~\ref{reshift} (b)).
On the other hand, the distances along $\pm(1,1,1)$-directions decrease 
while those along  $\pm(1,1,-1)$-directions increase for sites 3 and 4.
The changes of the other Ti-$R$ distances are rather small.
Since the nominal valence of $R$ ion is 3+, we expect that Ti $3d$ 
orbitals directed along the shorter Ti-$R$ bonds are lowered in energy
because of the attractive Coulomb interaction.
Without any distortions of the TiO$_6$ octahedra, the crystal field of the 
ligand oxygens has a cubic symmetry.
When we introduce the crystal field of $R$ cations 
($H_{R1}$) as a perturbation, the threefold degeneracy of the 
cubic-$t_{2g}$-levels split. 
Here, for simplicity, we take $H_{R1}$ by assuming that there exist the point 
charges with +3 valence on each $R$ cation.
The Coulomb interaction between an electron on a Ti $3d$ orbital 
and a ${\rm La}^{3+}$ ion is given by using the dielectric constant 
$\eps_{{\rm Ti}R}$ as,
\begin{equation}
 v({\vct r})=-\frac{Z_R e^2}{\eps_{{\rm Ti}R}|{\vct R}_i - {\vct r}|} ,
\end{equation}
where ${\vct R}_i$ expresses the coordinates of the $i$-th $R$ ion,
and $Z_R$(=+3) is the nominal valence of ${\rm La}^{+3}$ ion.
We calculate the following matrix elements,
\begin{equation}
 \langle m|v({\vct r})|m' \rangle= 
 \int {d{\vct r}} {{\varphi}^*}_{3d,m} v({\vct r}){\varphi_{3d,m'}}
\end{equation}
with 
\begin{equation}
 \varphi_{3d,m}=R_{3d}({\vct r}) Y_{2m}(\theta,\phi).  
\end{equation}
Here, the integer $m$(=$-2,...,2$) is the magnetic quantum number.
The coordinates of the La ions are calculated by using the 
positional parameters and the cell constants
obtained by the x-ray diffraction study~\cite{MacLean79}.
After transforming the basis from $m$ to the $t_{2g}$ representations, 
we obtain the Hamiltonian $H_{R1}$ with $xy$-, $yz$- and $zx$-basis.
These representations are defined in the coordinates with $x$-, $y$- and 
$z$-axes attached to each TiO$_6$ octahedron.

By diagonalizing the obtained $H_{R1}$, we can evaluate the $3d$-level 
energies and their representations.
The threefold degeneracy of the $t_{2g}$ orbitals
is split into three isolated levels as shown in Fig.~\ref{reshift} (c).
The energy levels $\Delta_1$ and $\Delta_2$ are 
0.7685/$\eps_{\rm TiLa}$ eV and 1.5849/$\eps_{\rm TiLa}$ eV, respectively. 
The representations of the lowest levels at each site can be specified 
by the linear combinations of $xy$, $yz$ and $zx$ orbitals as 
$axy-byz-czx$, $axy-cyz-bzx$, $axy+byz+czx$ and $axy+cyz+bzx$, respectively,
where $a^2$+$b^2$+$c^2$=1.
Here, we note that there exists a mirror plane vertical to the $c$-axis,
and the orbital structure has the same symmetry as the orthorhombic
GdFeO$_3$-type distortion.  
The coefficients take the values,
$a$=0.6046, $b$=0.3900 and $c$=0.6945.
This orbital structure is similar to that in the $D_{3d}$-crystal-fields 
with $[1,1,-1]$-, $[1,1,-1]$-, [1,1,1]- and [1,1,1]-trigonal axes
at sites 1, 2, 3 and 4, respectively. The representations of 
the lowest levels in these $D_{3d}$-crystal-fields are
$\frac{1}{\sqrt{3}}(xy-yz-zx)$, $\frac{1}{\sqrt{3}}(xy-yz-zx)$, 
$\frac{1}{\sqrt{3}}(xy+yz+zx)$ and $\frac{1}{\sqrt{3}}(xy+yz+zx)$,
respectively.
In the previous study~\cite{Mochizuki01b}, this configuration was referred 
to as config. 4, and it was shown that the $D_{3d}$-crystal-field 
with this trigonal-axes configuration strongly stabilizes the AFM(G) 
spin structure. 

We next discuss the stability of the magnetic 
state in the crystal field of $H_{R1}$.
For this purpose, we employ the effective spin and pseudospin 
Hamiltonian~\cite{Mochizuki00}: $H_{\rm eff.}=H_{\rm cry.}+H_{s\tau}$.
The first term $H_{\rm cry.}$ denotes the crystal field on the Ti $t_{2g}$ 
orbitals. The second term $H_{s\tau}$ is the spin and pseudospin term,
in which the threefold $t_{2g}$ orbital degrees of freedom
are represented by the pseudo-spin-1 operator $\tau$.
This effective Hamiltonian is derived from the multiband $d$-$p$ model
in the insulating limit by following an approach similar to the
Kugel-Khomskii formulation~\cite{Kugel72,Kugel73}.
In this $d$-$p$ model, the full degeneracies of Ti $3d$
and O $2p$ orbitals  as well as the on-site Coulomb and exchange
interactions are taken into account~\cite{Mochizuki01a}.
The on-site Coulomb interactions are expressed using Kanamori parameters 
$u$, $u^{\prime}$, $j$ and $j^{\prime}$ which
satisfy the following relations~\cite{Brandow77,Kanamori63}:
$u = U + \frac{20}{9}j$, $u'= u -2j$ and $j = j'$.
Here, $U$ gives the magnitude of the multiplet-averaged $d$-$d$ Coulomb 
interaction. The charge-transfer energy $\Delta$, which describes the 
energy difference between occupied O $2p$ and unoccupied Ti $3d$ levels, is 
defined using $U$ and the energies of the bare Ti $3d$ and O $2p$ orbitals 
$\eps_d^0$ and $\eps_p$ as $\Delta = \eps_{d}^0 + U -\eps_p$, since the 
characteristic unoccupied $3d$ level energy on the singly occupied Ti site 
is $\eps_{d}^0 + U$. The values of the parameters are estimated by the 
cluster-model analyses of valence-band and transition-metal $2p$ core-level 
photoemission spectra and the analyses of the first-principles band
calculations~\cite{Saitoh95,Mahadevan96}.
We take the values of these parameters as $\Delta = 5.5$ eV, $U = 4.0$ eV, 
$V_{pd\sigma} = -2.4$ eV, $V_{pd\pi} = 1.3$ eV, $V_{pp\sigma} = 0.52$ eV,
$V_{pp\pi} = -0.11$ eV and $j = 0.46$ eV. The effects of the 
${\rm GdFeO}_3$-type distortion are considered through the transfer 
integrals which are defined using Slater-Koster parameters~\cite{Harrison89}.
Substituting the $H_{\rm cry.}$ term with $H_{R1}$, we calculate the energies 
for several magnetic structures by applying the Hartree-Fock approximation. 
In Fig.~\ref{eneLaVMG}, we plot the calculated energies as functions of 
the magnitude of $\Delta_1$ (=$0.7685/{\eps}_{{\rm Ti}R}$).
We tune the magnitude of $\Delta_1$ by varying ${\eps}_{{\rm Ti}R}$.
In the region of $\Delta_1>$0.03 eV, the AFM(G) state is strongly 
stabilized relative to the other structures. 
\begin{figure}
$$ \psboxscaled{400}{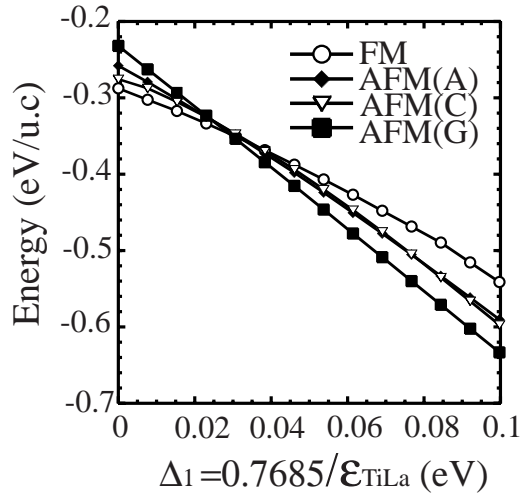} $$
\hfil
\caption{Energies for several magnetic structures under 
the crystal field of $H_{R1}$ are plotted as functions of 
$\Delta_1$.}
\label{eneLaVMG}
\end{figure}
In this region, $\Delta_1$ is much larger than $k_{\rm B}T_N$ 
so that the orbital occupation is restricted to the lowest level 
irrespective of the spin structure. We can estimate the 
magnitudes of the spin-exchange interactions for each Ti-Ti bond in the 
subspace of singly-occupied lowest levels. The spin-exchange constant $J$ 
can be represented as
$J=(E_{\uparrow\uparrow}-E_{\uparrow\downarrow})/2S^2$
with $E_{\uparrow\uparrow}$ and $E_{\uparrow\downarrow}$ being the energy 
gains of the Ti-Ti bond for $\uparrow\uparrow$- and  
$\uparrow\downarrow$-pairs, respectively.
For LaTiO$_3$, the values of $J$ along the $x$-, $y$- and $z$-axes
take as $J_x$=18.45 meV, $J_y$=18.45 meV and $J_z$=19.71 meV, respectively. 
They are in agreement with the value obtained by the neutron scattering 
experiment of $\sim$15.5 meV, and are consistent with the spin-wave 
spectrum well described by the isotropic Heisenberg model.

We further note that the crystal field of the La cations has two origins.
One is the Coulomb potential from charged La ions as we have studied above,
 and the other is the 
hybridization between the Ti $3d$ orbitals and unoccupied orbitals on the La 
cations. We next estimate the effect of the crystal field due to the latter 
origin for LaTiO$_3$. 
We construct the Hamiltonian for hybridization between Ti $3d$ and 
La $5d$ orbitals ($H_{R2}$) by using the second-order perturbational 
expansion in terms of the transfers between Ti $3d$ and La $5d$ orbitals 
($t^{dd}$). The expression of the matrix element of $H_{R2}$ is
\begin{equation}
 \langle m|H_{R2}|m' \rangle= -
 \sum_{i,\gamma} \frac{t_{m;i\gamma}^{dd}t_{m';i\gamma}^{dd}}{\Delta_{5d}}
\end{equation}
Here, the indices $m$ and $m'$ run over the cubic-$t_{2g}$-representations,
$xy$, $yz$ and $zx$. The symbols $i$ and $\gamma$ are indices for the eight 
nearest-neighbor La ions and the fivefold La $5d$ orbitals, respectively.
The symbol $\Delta_{5d}$ denotes the characteristic energy-difference
between Ti $t_{2g}$ and La $5d$ orbitals.
The transfers $t^{dd}$ are given in terms of Slater-Koster 
parameters $V_{dd{\sigma}}$, $V_{dd{\pi}}$ and $V_{dd{\delta}}$.
It is assumed that these parameters are proportional to $d^{-5}$
with $d$ being the Ti-La bond length.
On the basis of the analyses of LDA band structure, we fix these
parameters as $V_{dd{\sigma}}=-1.04$ eV, $V_{dd{\pi}}=0.56$ eV and 
$V_{dd{\delta}}=0$ eV for Ti-La bond length of 3.5 $\AA$,
and $\Delta_{5d}=3.6$ eV~\cite{Hamada02}.
By diagonalizing thus obtained $H_{R2}$, we have evaluated the 
energy-levels and their representations (see Table~\ref{tab:enelvl}). 
In the obtained energy-level structure, the threefold $t_{2g}$-levels
split into three nondegenerate levels, which is similar to
that of $H_{R1}$. Since the representation of the lowest level of $H_{R2}$ 
is also similar to that of $H_{R1}$, the level splitting of 
$H_{R1}$+$H_{R2}$ is well expressed by the sum of these two 
contributions. Considering the fact that AFM(G) state is stabilized in 
the region of $\Delta_1>$0.03 eV, and $\Delta_1$ for $H_{R2}$ is 
$\sim$0.085 eV, the crystal field due to the hybridization between Ti $3d$ 
and La $5d$ orbitals ($H_{R2}$) alone turns out to stabilize the AFM(G) 
spin structure strongly. In addition, the value of $\Delta_1$ is 
sufficiently large as compared with the coupling constant of the LS 
interaction in ${\rm Ti}^{3+}$ (${\zeta}_d$=0.018 eV)~\cite{Sugano70}.
Consequently, the crystal field of La cations dominates over
the LS interaction, resulting in the quenched orbital moment. 
The reduction of the ordered magnetic moment is attributed to 
the itinerant fluctuation instead of the LS interaction
near the metal-insulator phase boundary as discussed in
the literature~\cite{Mochizuki01b}.
\begin{table}[tdp]
\caption{The energy-level structures and the representations
of the lowest levels for $H_{R1}$ and $H_{R2}$.}
\label{tab:enelvl}
\begin{tabular} {lccccc}
 \hline
   \quad           & $H_{R1}$ & $H_{R2}$ &
   \quad           & $H_{R1}$ & $H_{R2}$ \\
 \hline
 \hline
  $\Delta_1$ (eV) :& 0.7685/$\eps_{\rm TiLa}$  & 0.0849 &  
  $a$ :            & 0.6046   & 0.6233 \\ 
  $\Delta_2$ (eV) :& 1.5849/$\eps_{\rm TiLa}$  & 0.1443 &  
  $b$ :            & 0.3900   & 0.4385 \\ 
  \quad            & \quad    & \quad                   &  
  $c$ :            & 0.6945   & 0.6474 \\   
  \hline
\end{tabular}
\end{table}

We have also studied the effects of the crystal field
of $R$ ions in $R$TiO$_3$ with $R$ being Pr, Nd and Sm.
The analyses of the crystal field 
Hamiltonian again reproduces the AFM(G) ground state in each compound.
Moreover, the gradual decrease of the spin-exchange constant $J$ as $R$ goes 
from La, Pr, Nd to Sm is reproduced (for example $J$ shows 20\% reduction at Nd
from La), which is consistent with the 
gradual decrease of $T_N$ of these compounds. The orbital 
states show that the occupation of the $yz$ orbital 
decreases at sites 1 and 3 as the GdFeO$_3$-type distortion increases 
while at sites 2 and 4, the $zx$-occupation decreases. 
As a result, the orbital structure in which sites 1, 2, 3 and 4 are 
occupied by $\frac{1}{\sqrt{2}}(xy-zx)$, 
$\frac{1}{\sqrt{2}}(xy-yz)$, $\frac{1}{\sqrt{2}}(xy+zx)$ and 
$\frac{1}{\sqrt{2}}(xy+yz)$, respectively is progressively stabilized. 
For example,
for Sm, the coefficients take $a=0.63, b=0.73$ and $c=0.24$
even without the JT distortion.
It has the 
similar symmetry as that realized in YTiO$_3$ with a large JT distortion, 
which is consistent with recent RXS study~\cite{Kubota00}. 
It is interesting to note that, in our results, the character of the orbital symmetry in YTiO$_3$ is retained but quantitatively and continuously decreases
toward the type of LaTiO$_3$ with the decreasing GdFeO$_3$-type distortion.
When the crystal field from O ion due to the JT distortion is superimposed, our calculation shows that the dominance 
is taken over by the JT mechanism at $R=$Sm.  Through severe competition at $R=$Sm, the Jahn-Teller 
mechanism comes to control the orbital-spin structure at $R=$Gd and Y.  

In summary, in order to elucidate the origin of the
AFM(G) state in RTiO$_3$ with $R$=Ti, we have examined the crystal field of 
La cations by considering the experimentally observed distortion of the La ions.
The crystal field Hamiltonian is constructed
by using the experimental position parameters.
Based on this Hamiltonian, we have shown
that the distortion of the La ions caused by the GdFeO$_3$-type distortion generate 
the crystal field with lower symmetry, which is similar to the $D_{3d}$ crystal field.
Then the threefold degenerate cubic-$t_{2g}$-levels
split into three isolated levels. 
The energies and the spin-exchange constant
calculated by the effective Hamiltonian
show that the lowest-level occupations in this crystal field
stabilize the AFM(G) state, and account for the spin-wave spectrum 
for LaTiO$_3$ with (1) $J \sim 15.5$ meV,
(2) isotropic spin-coupling, and (3) quenched orbital moment.
The orbital-spin structre for $R=$ Pr, Nd and Sm is also accounted for by the same mechanism.   
GdFeO$_3$-type distortion has a universal relevance in determining
the orbital-spin structure of perovskite compounds in competition with the JT mechanism. 

We thank N. Hamada for providing us with data on LDA calculations of
LaTiO$_3$.
This work is supported by ``Research for the Future Program''
(JSPS-RFTF97P01103) from the Japan Society for the Promotion of Science.

\end{document}